\documentclass [12pt,a4paper]{article}
\usepackage{bbm}
\usepackage{amsfonts}
\usepackage{mathrsfs}
\usepackage {amssymb}
\usepackage {amsmath}
\usepackage{amsthm}
\usepackage{latexsym}
\usepackage {amssymb}
\usepackage {amsmath}
\usepackage{color}

\theoremstyle{definition}

\newtheorem{theorem}{Theorem}

\newtheorem{lemma}{Lemma}

\newcommand{\F}{\mathbb{F}}

\makeatletter
 \@addtoreset{equation}{section}
 
\makeatother

\usepackage{lastpage}
\usepackage{epsfig}

\newcommand{\pf}{{\bf Proof. \ }}
\begin{document}
\begin{center}
\textbf{\Large{The $(n,m,k,\lambda)$-Strong External Difference Family with $m \geq 5$ Exists } }\footnote { The work of J. Wen and M. Yang is supported by the NSFC under Grant 61171082, 61379139. The work of K. Feng was supported by the NSFC under Grant 11471178, 11571107.\\
J. Wen is with the Chern Institute of Mathematics, Nankai University, Tianjin 300071, China(e-mail:jjwen@mail.nankai.edu.cn)\\
 M. Yang is with the State Key Laboratory of Information
Security, Institute of Information Engineering, Chinese Academy
of Sciences, Beijing 100195, China(e-mail:yangminghui6688@163.com)\\
K. Feng is with the Department of Mathematical Sciences, Tsinghua University, Beijing, 100084, China(e-mail: kfeng@math.tsinghua.edu.cn)}

\end{center}

\begin{center}
\small Jiejing Wen, Minghui Yang and  Keqin Feng
\end{center}

%-------------------------------------------------------------------------

\noindent\textbf{Abstract} The notion of strong external difference family (SEDF) in a finite abelian group $(G,+)$ is
raised by M. B. Paterson and D. R. Stinson [5] in 2016 and motivated by its application in communication theory to construct
$R$-optimal regular algebraic manipulation detection code. A series of $(n,m,k,\lambda)$-SEDF's have been constructed in [5, 4, 2, 1]
with $m=2$. In this note we present an example of (243, 11, 22, 20)-SEDF in finite field $\mathbb{F}_q$ $(q=3^5=243).$ This is an
answer for the following problem raised in [5] and continuously asked in [4, 2, 1]: if there exists an $(n,m,k,\lambda)$-SEDF for
$m\geq 5$.

\section{Introduction}
\ \ \ \ Let $(G,+)$ be a finite abelian group with $n=|G|$ elements. For subsets $D_1$ and $D_2$ of
$G$ $(|D_1|,|D_2|\geq 1)$, we defined the following multiset
$$\Delta(D_1, D_2)=\{a_1-a_2: a_1\in D_1, a_2\in D_2\}$$

$\mathbf{Definition}$ Let $A_1,\ldots,A_m$ $(m\geq 2)$ be subsets of $G$, $|A_i|=k$ $(1\leq i\leq m)$.
The family $\{A_1,\ldots,A_m\}$ is called an $(n, m, k, \lambda)$-strong external difference family (SEDF)
in $G$ if for each $i$ $(1\leq i\leq m)$,
$$\sum^{m}_{j=1\atop j\neq i }\Delta(D_i, D_j)=\lambda(G-\{0\})\ \ \ \ \ \ \ \ \ \ (1)$$
where we use the notation  in the group ring $\mathbb{Z}[G]$ to express the multisets in both sides of equality (1).
Namely, this equality means that for each nonzero element $g$ in $G$, the multiplicity of $g$ in the multiset
$$\bigcup^{m}_{j=1\atop j\neq i }\Delta(D_i, D_j)$$ is constant $\lambda$, and 0 is not in this multiset. It is
easy to see that the equality (1) implies that $(m-1)k^2=\lambda(n-1)$ and $A_1,\ldots,A_m$ should be pairwise disjoint.

Let $G=\{g_1,\ldots,g_n\}$. Then  $\{\{g_1\},\ldots,\{g_n\}\}$ is the trivial $(n,n,1,1)$-SEDF in $G$. From now on,
we concern on nontrivial SEDF's.

The notion of SEDF (and its generalizations) is raised by M. B. Paterson and D. R. Stinson [5] and motivated
by its application in communication theory to construct $R$-optimal regular algebraic manipulation detection
codes. Many series of $(n,m,k,\lambda)$-SEDF's  with $m=2$ have been constructed in [5, 4, 2, 1]. Paterson and
Stinson [5] raised a problem on existence of SEDF with $m\geq 3$. Later, Martin and Stinson [4] proved that
there is no (notrivial) SEDF with $m=3$ and 4. Then Haczunska and Paterson [2] presented more nonexistence results
and asked if there exists an SEDF with $m\geq 5$. Very recently, J. Bao et al. [1]  presented even more nonexsitence
results and conjectured that such SEDF does not exist.

In this note we show an example of SEDF with parameters $(n,m,k,\lambda)=(243,11,22,\\20)$ which gives an
answer of above mentioned problem. More precisely, we will show that the cyclotomic classes
$\{C_0,C_1,\ldots,C_{10}\}$ of order 11 in finite field $\mathbb{F}_q$ $(q=3^5=243)$ is a
(243, 11, 22, 20)-SEDF in $(\mathbb{F}_q,+)$. Since the example is very concrete, our presentation
and computation are down to the earth. Firstly, we determine a primitive element $\theta$ of $\mathbb{F}_q$,
$\mathbb{F}_q^{\ast}=\langle\theta\rangle$. For doing this, we consider the polynomial
$$f(x)=x^5+x^4+x^3+x^2+2x+1\in \mathbb{F}_3[x].$$
From $g(0)=g(1)=1$ and $g(2)=g(-1)=-1$ we know that $f(x)$ has no factor $x-a$ $(a\in \mathbb{F}_3$). Moreover,
$f(x)$ is not divided by the quadratic irreducible polynomials $x^2+1$ and $x^2\pm x-1$ in  $\mathbb{F}_3[x]$.
Therefore $f(x)$ is irreducible in $\mathbb{F}_3[x]$ and $\mathbb{F}_q=\mathbb{F}_3(\theta)$ where $\theta$
is a root of $f(x).$

\begin{lemma} $\theta$ is a primitive element of $\mathbb{F}_q$ $(q=3^5).$
\end{lemma}
\pf{ Since $\{1, \theta, \theta^2, \theta^3, \theta^4\}$ is a basis of $\mathbb{F}_q$ over $\mathbb{F}_3$,
each element $\alpha$ can be expressed uniquely as
$$\alpha=c_0+c_1\theta+c_2\theta^2+c_3\theta^3+c_4\theta^4   \  \ (c_i\in \mathbb{F}_3)$$
We identify $\alpha$ as the vector $(c_0, c_1, c_2, c_3, c_4)$ in $\mathbb{F}_3^{5}$ and denote
$\alpha=(c_0 c_1 c_2 c_3 c_4)$ briefly. Therefore\\
$\theta^{0}=1=(10000), \theta=(01000), \theta^2=(00100),  \theta^3=(00010), \theta^4=(00001)$
and $\theta^{5}=(21222)$ since $0=f(\theta)=\theta^5+\theta^4+\theta^3+\theta^2+2\theta+1$.

In general, if $\theta^t=(c_0c_1c_2c_3c_4)$, then
$$\theta^{t+1}=(0 c_0 c_1 c_2 c_3)+c_4\theta^5=(0c_0c_1c_2c_3)+c_4(21222).$$
From this recursive formula, we can get \\
$$\theta^6=(11200),\theta^7=(01120),\theta^8=(00112),\theta^9=(12122),\theta^{10}=(10020),\theta^{11}=(01002)$$
and $\theta^{12}=(12211).$
We need to show that the (multiplicative) order of $\theta$ is $q-1=242=2\cdot 11^{2}$. For doing this we
need to show that $\theta^{22}\neq 1$ and $\theta^{121}\neq 1$. Firstly,
\begin{equation*}\begin{split}
\theta^{22}& =(\theta^{11})^2=(\theta+2\theta^4)^2=\theta^2+\theta^5+\theta^8=(00100)+(21222)+(00112)\\
                                  &= (21101)\neq 1.
\end{split}\end{equation*}
Next
$$\theta^{33}={(\theta^{11})}^3=(\theta+2\theta^4)^3=\theta^3+2\theta^{12}=(00010)+(21122)=(21102)$$
$$\theta^{99}={(\theta^{33})}^3=(2+\theta+\theta^2+2\theta^4)^3=2+\theta^3+\theta^6+2\theta^{12}=(22002)$$
and
\begin{equation*}\begin{split}
\theta^{121}& =\theta^{22}\cdot\theta^{99}=(2+\theta+\theta^2+\theta^4)(2+2\theta+2\theta^4)=1+\theta^2+2\theta^3+\theta^5+2\theta^6+2\theta^8\\
                                  &= (20000)=2\neq 1.
\end{split}\end{equation*}
Therefore $\theta$ is a primitive element of $\mathbb{F}_q$ $(q=3^5).$} \qed

Next, we need a little knowledge on cyclotomic classes in finite field and cyclotomic numbers. Let $q=p^m$ where $m\geq1$
and $p$ is a prime, $q-1=ef$ $(e\geq 2)$, $\mathbb{F}_q^{\ast}=\langle\theta\rangle$. Then $C=\langle\theta^e\rangle$ is the
cyclic subgroup of $\mathbb{F}_q^{\ast}$ with size $|C|=f$. All cosets $C_\lambda=\theta^{\lambda}C$ $(0\leq\lambda\leq e-1)$
of $C$ in $\mathbb{F}_q^{\ast}$ are called the cyclotomic classes of order $e$. The cyclotomic numbers $(i,j)_e$ $(0\leq i,j\leq e-1)$
over $\mathbb{F}_q$ are defined by
$$(i,j)_e=|(1+C_i)\cap C_j|=\sharp\{x\in C_i: 1+x\in C_j\}.$$

The following properties of $(i,j)_e$ can be seen in T. Storer's book [6].

\begin{lemma} Let $q=p^m$ where $m\geq 1$ and $p$ is an odd prime, $q-1=ef$ $(e\geq 2), C_\lambda$ $(0\leq\lambda\leq e-1)$
be the cyclotomic classes of order $e$ in $\mathbb{F}_q,$ $(i,j)_e$ $(0\leq i,j\leq e-1)$ be the cyclotomic numbers over $\mathbb{F}_q$.
Then

(1) For $r, s, r', s' \in \mathbb{Z}$ and $r\equiv r'(\bmod e),$ $s\equiv s'(\bmod e),$ we have \\
 $$ C_r=C_{r'} \ \text{and} \ (r,s)_e=(r',s')_e.$$

(2)$(i,j)_e=(-i, j-i)_e$ and $(i,j)_e=(pi, pj)_e$ $(0\leq i,j\leq e-1).$

(3)$\Delta(C_0,C_0)=f\cdot\{0\}+\sum_{\lambda=0}^{e-1}(e-\lambda, e-\lambda)_eC_{\lambda}.$

(4)If $2|f$, then $-1\in C_0,$ $-C_\lambda=C_\lambda,$ $0\leq\lambda\leq e-1)$ and $(i,j)_e=(j,i)_e$.
\end{lemma}
In this paper we need to compute $(i,i)_{11}\ (0\leq i \leq 10)$ over $\F_q\ (q=3^5)$.
\begin{lemma}
Let $q=3^5=243, q-1=ef$ where $e=11$ and $f=22$, $(i,j)=(i,j)_{11},\ (0\leq i,j\leq 10)$
be the cyclotomic number of order $e=11$ over $\F_q$. Then $(0,0)=1$ and
$(i,i)=2$ for all $1\leq i\leq 10$.
\end{lemma}
\pf{
Let $\F_q^*=\langle \theta \rangle$ where $\theta$ is the primitive element of $\F_q$
given in Lemma~1. We have the formula
\[
\begin{aligned}
\Delta(C_0,C_0)&= f\{0\}+\sum_{\lambda=0}^{10}(11-\lambda,11-\lambda)C_{\lambda}\   (\text{Lemma}\ 2(3)) \\
               &= f\{0\}+\sum_{\lambda=0}^{10}(\lambda,\lambda)C_{11-\lambda}
\end{aligned}
\]
which implies that $f^2=f+f\sum_{\lambda=0}^{10}(\lambda,\lambda)$. Namely,
$\sum_{\lambda=0}^{10}(\lambda,\lambda)=f-1=21$. By $(i,i)=(3i,3i)$ (Lemma 2 (2)), we get
\[
(1,1)=(3,3)=(9,9)=(5,5)=(4,4),(2,2)=(6,6)=(7,7)=(10,10)=(8,8).
\]
Let $A=(0,0),B=(9,9)$ and $C=(7,7)$. Then
$$
21=\sum_{\lambda=0}^{10}(\lambda,\lambda)=A+5(B+C)\ \ \ \ \ \ \ \  \ \ \ (2)
$$
Now we compute the set $C_0=\langle \theta^{11} \rangle=\{\theta^{11\cdot l}: 0\leq l\leq21\}$
explicitly. From $-1=\theta^{\frac{q-1}{2}}=\theta^{121}\in C_0$ we know that
\[
C_0=D+(-D),\   D=\{\theta^{11\cdot l}: 0\leq l\leq 10\}.
\]
We have computed in the proof of Lemma~1 that
\[
\theta^{11}=(01002),\theta^{22}=(21101),\theta^{33}=(21102),\theta^{99}=(22002).
\]
With the same way of computation we get (by $\theta^{121}=-1$)
\[
\theta^{55}= (\theta^{99})^3 =(11112),\theta^{44}=-(\theta^{55})^3=(12212),\theta^{66}=(\theta^{22})^3=(10121),
\]
\[
\theta^{77}=-(\theta^{66})^3 =(12011),\theta^{110}=-(\theta^{77})^3=(01020),\theta^{88}=(\theta^{110})^3=(12112).
\]
Namely, $C_0=D+(-D)$ and
\[
\begin{aligned}
D=&\{\theta^{11\cdot l}: 0\leq l\leq 10\}  \\
 =&\{(10000),(01002),(21101),(21102),(12212),(11112), \\
  &  (10121),(12011),(12112),(22002),(02010)\}.
\end{aligned}
\]
From $\pm 1\in C_0$ and $(-1)-1=1$ we know that $A=(0,0)\geq 1$.
From $\theta^{22}=(21101)\in D\subseteq C_0$, $\theta^{33}=(21102)\in D\subseteq C_0$
and $\theta^{33}-\theta^{22}=(00001)=\theta^{4}$ we know that
$\pm\alpha=\pm \theta^{33-4}\in C_7$, $\pm\beta =\pm \theta^{22-4}\in C_7$, $\alpha-\beta=1$,
$-\beta-(-\alpha)=1$ and $\alpha=\theta^{29}\neq -\beta=-\theta^{18}$.
Therefore $C=(7,7)\geq 2$. Similarly, from $\theta^{44}=(12212)\in D, \theta^{88}=(12112)\in D$
and $\theta^{44}-\theta^{88}=\theta^2$ we know that $\pm \gamma=\pm \theta^{44-2}\in C_9$,
$\pm\delta=\pm \theta^{88-2}\in C_9$, $\gamma-\delta=(-\delta)-(-\gamma)=1$
and $\gamma=\theta^{42}\neq -\theta^{86}=-\delta$. Therefore $B=(9,9)\geq 2$.
Then by~(2) we get
\[
21=A+5(B+C)\geq 1+5(2+2)=21,
\]
which implies that $A=1$ and $B=C=2$. Namely, $(0,0)=1$ and $(i,i)=2$
for all $1\leq i\leq 10$. This completes the proof of Lemma 3.
\qed
}

Now we come to the main result.
\begin{theorem}
Let $q=3^5=243$, $C_{\lambda} (0\leq \lambda \leq 10)$ be the cyclotomic classes of order
$11$ of $\F_q$. Then $\{C_0,C_1,\ldots, C_{10}\}$ is an $(n,m,k,\lambda)=(243,11,22,20)$-SEDF
in $(\F_q,+)$.
\end{theorem}
\pf{
From $\F_q=\{0\}+\sum_{\lambda=0}^{10}C_{\lambda}$ we get that for each $i\ (0\leq i\leq 10)$,
\[
\begin{aligned}
\sum_{\substack{{j=0}\\ {j\neq i}}}^{10}\Delta(C_i,C_j)&=\Delta(C_i,\sum_{\substack{{j=0}\\ {j\neq i}}}^{10}C_j)=\Delta(C_i,\F_q-C_i-\{0\}) \\
&=f\F_q - \Delta(C_i,C_i)-C_i\ (\text{for any}\ S\subseteq \F_q,\ \Delta(S,\F_q)=|S|\cdot \F_q) \\
&=f(\F_q-\{0\})-\sum_{\lambda=0}^{10}(i-\lambda,i-\lambda)C_{\lambda}-C_i \\
&=f(\F_q-\{0\})-(C_i+2\sum_{\substack{{\lambda=0}\\ {\lambda\neq i}}}^{10}C_{\lambda})-C_i\  (\text{by Lemma~3})\\
&=f(\F_q-\{0\})-2\sum_{\lambda=0}^{10}C_{\lambda} \\
&=(f-2)(\F_q-\{0\}) \\
&=20(\F_q-\{0\}),
\end{aligned}
\]
which means that $\{C_0,C_1,\ldots,C_{10}\}$ is a $(243,11,22,20)$-SEDF in $\F_q\ (q=3^5)$.
\qed
}

$\mathbf{Remark}$
Our example is not a coincidence. From
\[
\Delta(C_i,C_i)=2\sum_{\substack{{j=0}\\ {j\neq i}}}^{10}C_j+C_i+f\{0\}
=2(\F_q-C_i-\{0\})+C_i+22\{0\}
\]
we know that each $C_i\ (0\leq i\leq 10)$ is an $(n,k,\lambda,\mu)$-partial difference set
(PDS) in $\F_q\ (q=3^5)$ where $n=q,k=|C_i|=22,\mu=2$ and $\lambda=1=\mu-1$.
It is shown~([5], Theorem 2.4) that if $A_1,\ldots,A_m$ is a partition of
an abelian group $G$ and each $A_i$ is a $(n,k,\lambda)$-difference set in $G$,
then $\{A_1,\ldots,A_m\}$ is an $(n,m,k,\lambda')$-SEDF in $G$ with $\lambda'=k-\lambda$.
Similarly we can show that if $A_1,\ldots,A_m$ is a partition of $G-\{0\}$ and each
$A_i$ is an $(n,k,\lambda,\mu)$-PDS in $G$ with $\lambda=\mu-1$, then $\{A_1,\ldots,A_n\}$
is an $(n,m,k,\lambda')$-SEDF in $G$ with $\lambda'=k-\lambda$. In sequential paper
we will use the PDS with $\lambda=\mu-1$ to construct SEDF with $m=2$,
Generalized SEDF and Bounded GSEDF. (For the definition of GSEDF and BGSEDF and
their applications we refer to [5]).

On the other hand, it is proved that any nontrivial PDS with $\lambda=\mu-1$
and $k<\frac{n}{2}$ in a finite abelian group has only two types of parameters
$(n,k,\lambda,\mu)=(n,\frac{n-1}{2},\frac{n-5}{4},\frac{n-1}{4})$
and $(243,22,1,2)$. The possible SEDF constructed by using these two types
of PDS has parameters $m=\frac{n-1}{k}=2$ and $11$. At this moment we still
wonder if there exists $(n,m,k,\lambda)$-SEDF with $m\geq 5$ and $m\neq 11$.

\end{document}